\documentclass[twocolumn,aps,prl,showpacs,superscriptaddress]{revtex4-2}
\usepackage[T1]{fontenc}
\usepackage{babel}
\usepackage{amsmath}
\usepackage{amssymb}
\usepackage{graphicx}
\usepackage{wasysym}
\usepackage[unicode=true]
 {hyperref}

\makeatletter
\renewcommand{\fnum@figure}{FIG.~\thefigure}

\usepackage{algorithm,algpseudocode}
\usepackage{color,xcolor}
\usepackage[caption=false]{subfig}
\definecolor{darkRed}{RGB}{220,0,0}
\definecolor{darkGreen}{RGB}{0,130,0}
\definecolor{darkBlue}{RGB}{0,0,220}
\usepackage{hyperref}
\hypersetup{
      colorlinks=true,
      citecolor=darkGreen,
      linkcolor=darkGreen,
      urlcolor=darkGreen}
\makeatother

\begin{document}

\title{Non-Hermitian skin effects on many-body localized and thermal phases}

\author{Yi-Cheng Wang}
\thanks{These authors contributed equally to this work.}
\affiliation{Institute of Atomic and Molecular Sciences, Academia Sinica, Taipei 10617, Taiwan}
\affiliation{Department of Physics, National Taiwan University, Taipei 10617, Taiwan}
\author{Kuldeep Suthar}
\thanks{These authors contributed equally to this work.}
\affiliation{Institute of Atomic and Molecular Sciences, Academia Sinica, Taipei 10617, Taiwan}
\author{H. H. Jen}
\affiliation{Institute of Atomic and Molecular Sciences, Academia Sinica, Taipei 10617, Taiwan}
\affiliation{Physics Division, National Center for Theoretical Sciences, Taipei 10617, Taiwan}
\author{Yi-Ting Hsu}
\email{yhsu2@nd.edu}
\affiliation{Department of Physics, University of Notre Dame, Notre Dame, Indiana 46556, USA}
\author{Jhih-Shih You}
\email{jhihshihyou@ntnu.edu.tw}
\affiliation{Department of Physics, National Taiwan Normal University, Taipei 11677, Taiwan}

\date{\today}

\begin{abstract}
Localization in one-dimensional interacting systems can be caused by disorder potentials or non-Hermiticity.
The former phenomenon is the many-body localization (MBL), and the latter is the many-body non-Hermitian skin effect (NHSE).
In this work, we numerically investigate the interplay between these two kinds of localization, where the energy-resolved MBL arises from a deterministic quasiperiodic potential in a fermionic chain.
We propose a set of eigenstate properties and long-time dynamics that can collectively distinguish the two localization mechanisms in the presence of non-Hermiticity.
By computing the proposed diagnostics, we show that the thermal states are vulnerable to the many-body NHSE while the MBL states remain resilient up to a strong non-Hermiticity.
Finally, we discuss experimental observables that can probe the difference between the two localizations in a non-Hermitian quasiperiodic fermionic chain.
Our results pave the way toward experimental observations on the interplay of interaction, quasiperiodic potential, and non-Hermiticity.
\end{abstract}

\maketitle

{\textit{Introduction.}}---Many-body localization~(MBL)~\cite{Nandkishore2015,Alet2018,Abanin2019} can exist in one-dimensional (1D) isolated quantum systems in the presence of interaction and disorders, where thermalization fails to occur~\cite{Basko2007,Oganesyan2007,Pal2010,Serbyn2013} and the information encoded in the initial state is preserved~\cite{Schreiber2015,Choi2016,Smith2016}.
Besides the well-known cases with random disorders~\cite{Gornyi2005,Basko2006,Oganesyan2007,Znidaric2008,Pal2010,Kjall2014,Luitz2015}, where the thermal to MBL transition occurs as the disorder strength increases, numerical~\cite{Iyer2013,Li2015,Modak2015,Li2016,Hsu2018,Khemani2017,Zhang2018,Znidaric2018,Doggen2019,Xu2019,An2021,WangYunfei2022} and experimental~\cite{Schreiber2015,Roushan2017,Kohlert2019} evidences have suggested that MBL can also occur in the presence of deterministic but quasiperiodic potentials.
In particular, in quasiperiodic systems with a single-particle mobility edge, MBL and thermal phases have been found to coexist at a given intermediate potential strength in low- and mid-spectrum regimes, respectively~\cite{Li2015,Modak2015,Li2016,Hsu2018}.
Such an energy-resolved localization-delocalization transition is originated from the non-trivial interplay between the interaction and quasiperiodic potential, where the latter provides a localization mechanism in many-body Hermitian systems.

In non-Hermitian systems, a distinct localization mechanism dubbed non-Hermitian skin effect~(NHSE) has recently attracted rapidly growing theoretical~\cite{Yao2018,Martinez2018,Kunst2018,Yokomizo2019,Lee2019,Longhi2019,Borgnia2020,Okuma2020,Zhang2020,Kawabata2020,Scheibner2020,Yi2020,Li2020,Zhang2022,Wang2022} and experimental~\cite{Helbig2020,Hofmann2020,Ghatak2020,Weidemann2020,Xiao2020,Xiao2021,Zhang2021,Lin2022NC,Liang2022,Lin2022} attention, where an extensive number of eigenstates are localized at open boundaries.
In the non-interacting limit, single-particle NHSE has been shown to occur under open boundary condition when the eigenspectrum under periodic boundary condition exhibits nontrivial winding~\cite{Okuma2020,Zhang2020}. This can be viewed as the non-Hermitian analogue of the ``bulk-boundary correspondence'' in topological systems. In terms of specific models, it has been shown that such winding and localization can arise when the non-Hermiticity is introduced by nonreciprocal hoppings~\cite{Hatano1996}. 
In the presence of interactions, although the relation between the winding in eigenspectrum and many-body NHSE remains an interesting but elusive topic, recent theoretical works have investigated the existence~\cite{Mu2020,Alsallom2022} and entanglement dynamics~\cite{Kawabata2022} of many-body NHSE in fermionic systems, how MBL is affected by non-Hermiticity \cite{Hamazaki2019,Zhai2020}, as well as NHSE in random disordered systems \cite{Kuldeep2022}.
Although the many-body NHSE does not exhibit strictly exponential localization in the real space as the single-particle NHSE does due to Pauli exclusion principle~\cite{Alsallom2022}, particles in all the many-body eigenstates are still expected to accumulate on one end of open boundaries in the strong non-Hermiticity limit.
Therefore, in stark contrast to MBL, where particles in a given initial states stay localized at their initial positions, many-body NHSE tends to push all particles towards one of the open boundaries such that the initial information is lost. Yet, the competition between these two localizations in interacting 1D systems with both non-Hermiticity and quasiperiodic potentials remains elusive. 

In this Letter, we investigate how many-body NHSE affects the MBL and thermal phases in 1D quasiperiodic systems, focusing on a case study of the generalized Aubry-Andr\'e~(GAA) model~\cite{Ganeshan2015,Li2015,Modak2015,Zeng2020} in the presence of nonreciprocal hoppings.
To distinguish the two localization mechanisms, namely the quasiperiodic potential and the non-Hermiticity, we propose a set of eigenstate properties and dynamical responses that can collectively diagnose the many-body NHSE and the non-Hermitian localized phase connected to the Hermitian MBL, dubbed non-Hermitian MBL.
Our key finding is that the thermal phase is vulnerable to non-Hermiticity such that the volume-law entanglement entropy vanishes and the many-body NHSE appears already at a small asymmetry between the left and right hopping.
In contrast to the thermal phase, MBL dominates over the many-body NHSE up to an extremely large hopping asymmetry, where the many-body eigenstates across the full spectrum exhibit strong NHSE.
Importantly, we emphasize that although both quasiperiodic potentials and non-Hermiticity can drive localization, the many-body NHSE does \textit{not} preserve the information of the initial state as does the MBL phase.
Finally, we discuss possible experimental detection scheme that distinguishes non-Hermitian MBL and many-body NHSE using the long-time behaviors of generalized imbalance~\cite{G2021,Guo2021,Morong2021} and local particle numbers. 
We expect that our observation of how the many-body NHSE affects the thermal and MBL phases is generic for other non-Hermitian systems with many-body NHSE in the presence of quasiperiodic as well as random disorder potentials.

\begin{figure}[!t]
\centering{}
\includegraphics{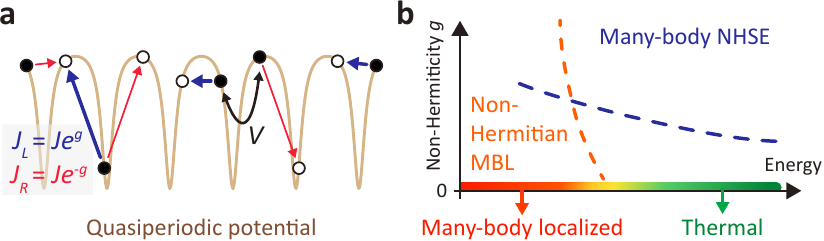}
\caption{\label{fig:1}
(a)~Schematic of an open fermionic chain subject to generalized Aubry-Andr\'e potential~(brown, $\phi=0$).
The nonreciprocal hoppings $J_{L(R)}=Je^{\pm g}$~(blue and red) and nearest-neighbor interaction $V$~(black) are shown according to the configuration of occupied~($\bullet$) and unoccupied~($\circ$) sites.
(b)~The phase diagram as a function of energy and non-Hermiticity parameter $g$ explored in this Letter.
The red and blue dashed lines roughly separate the parameter regimes with the initial state memory and localization of particles at the open boundary, respectively.
}
\end{figure}

{\textit{Model.}}---In the absence of interactions, models with non-reciprocal hoppings are well-known to exhibit winding in the eigenspectrum and thus single-particle NHSE, and could realized experimentally in various systems~\cite{Su2022,Zhang2021a,Gong2018}. Now in the presence of interactions, we therefore investigate many-body NHSE by introducing non-Hermiticity with non-reciprocal hoppings. Specifically, we consider a non-Hermitian interacting GAA model that has a non-reciprocal factor $e^{\pm g}$ in the nearest-neighbor hopping terms~[Fig.~\ref{fig:1}(a)]
\begin{align}
H= & \sum_{j=1}^{L-1}\Big[-J\big(e^{g}c_j^\dagger c_{j+1}+e^{-g}c_{j+1}^\dagger c_j\big)+V n_j n_{j+1}\Big]\nonumber
\\
+ & \sum_{j=1}^{L}2\lambda\frac{\displaystyle\cos(2\pi q j+\phi)}{\displaystyle 1-\alpha\cos(2\pi q j+\phi)}n_j,\label{eq:1}
\end{align}
where $c_j^\dagger$ creates a fermion on site $j$ in an open chain with $L$ sites, $n_j=c_j^\dagger c_j$ is the density operator, $V$ is the nearest-neighbor density-density interaction, and $g\neq0$ controls the strength of non-Hermiticity.
The model has a quasiperiodic potential with strength $2\lambda$, an irrational wave number $q=(\sqrt{5}-1)/2$, a randomly chosen global phase $\phi$, and a dimensionless parameter $\alpha=-0.8$ that controls the potential shape.
In the noninteracting limit, this model exhibits an exact single-particle mobility edge at energy $E=2\text{sgn}(\lambda)(|J|-|\lambda|)/\alpha$~\cite{Ganeshan2015}.
At nonzero interaction, previous numerical studies~\cite{Li2015,Hsu2018} on a Hermitian GAA model~($g=0$) with a moderate potential strength found MBL and thermal states in the low- and mid-spectrum regimes, respectively, along with an intermediate phase between them~[Fig.~\ref{fig:1}(b)].
In the following, we choose $V/J=1$ and $\lambda/J=0.45$ such that both MBL and thermal states exist in the low- and mid-spectrum regimes in the Hermitian limit $g=0$, respectively. 

{\textit{Non-Hermiticity induced localization: NHSE.}}---To investigate the role of non-Hermiticity in an interacting quasiperiodic system, we first note that the eigenspectrum of Eq.~(\ref{eq:1}) is independent of $g$ regardless of total particle number.
This is because it can be mapped from a Hermitian GAA model by the imaginary gauge transformation~\cite{Hatano1996}, $c_j\to e^{gj}c_j$ and $c_j^\dagger\to e^{-gj}c_j^\dagger$.
As a result, the level spacing statistics, a commonly used diagnostic that indicates the MBL and thermal phases in Hermitian system~\cite{Oganesyan2007,Iyer2013}, cannot reflect the occurrence of NHSE or the fate of MBL and thermal states at $g\neq0$~\cite{SupplementaryMaterial}.
Nevertheless, we expect that the influence of NHSE can be identified from eigenstate-related quantities since the probability of many-body product states $|n\rangle$ with particles accumulated near one of the open ends is enhanced as $|n\rangle\to e^{-g\sum_{j=1}^{L}j n_j}|n\rangle$ followed by normalization. 
We therefore investigate two eigenstate properties in the following to capture the impact of NHSE.

The two eigenstate-based quantities we study are the energy-resolved real-space local density $\langle n_j\rangle$ and R\'enyi entropies $S_2(l,L)=-\log[\text{Tr}\rho_A^2]$, where $\rho_A=\text{Tr}_B\big[|\psi\rangle\langle\psi|\big]$ is the reduced density matrix of eigenstate $|\psi\rangle$ for the left subsystem $A$ that consists of sites $i=1,2,...,l\leq L$ in an $L$-site open chain, obtained by tracing out the right subsystem $B$. 
Both energy-resolved quantities are computed by eigenstates uniformly sampled throughout the spectrum using exact diagonalization (ED) for an $L=30$ chain with $N=5$ particles. 
Since the eigen-energies remain unchanged under the non-Hermiticity magnitude $g$, we can study how $\langle n_j\rangle$ and $S_2(l,L)$ at different energies change from the Hermitian $g=0$ to the non-Hermitian $g>0$ cases.

\begin{figure}[!t]
\centering{}
\includegraphics{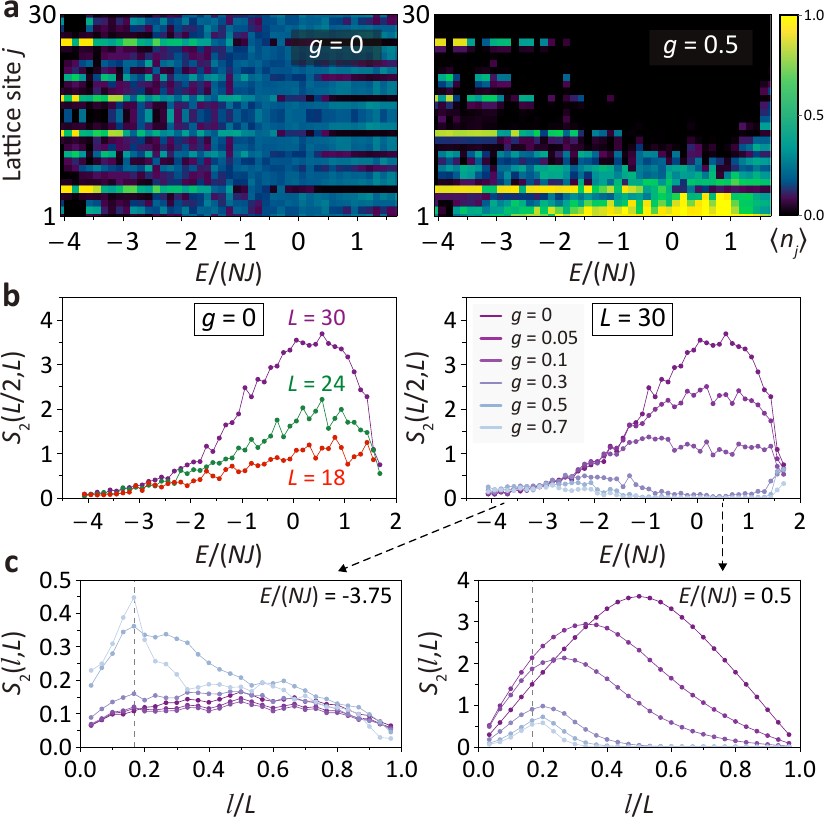}
\caption{\label{fig:2}
(a)~Energy-resolved local particle numbers in real space for Hermitian~(left, $g=0$) and non-Hermitian~(right, $g=0.5$) cases.
The latter shows localization at the left end induced by NHSE in mid-spectrum regime.
(b)~In Hermitian case~(left), eigenstates exhibit area-law and volume-law entanglement scalings in the low- and mid-spectrum regimes, respectively.
The nonreciprocal hoppings~(right) result in the entanglement reduction in the mid-spectrum regime.
(c)~R\'enyi entropy as a function of subsystem size $l$ at low-~(left, MBL at $g=0$) and mid-~(right, thermal at $g=0$) spectrum regimes.
The color gradient follows that in the right panel of (b), and the dashed vertical line indicates $l/L=N/L$.
The plots are obtained by averaging over $\phi$~\cite{SupplementaryMaterial}, and the filling $N/L=1/6$ is considered throughout this work.
}
\end{figure}

We first show the drastic energy dependence in how the real-space local density $\langle n_j\rangle$ evolves with the non-Hermiticity $g$ (see Fig.~\ref{fig:2}(a)). 
At $g=0$ in the left panel, the low- and mid-spectrum eigenstates show symmetric $\langle n_j\rangle$ with respect to the center of 1D chain, such that any asymmetric local density at nonzero $g$ reflects the influence of NHSE.
As we turn on the non-Hermiticity at $g=0.5$ in the right panel, the mid-spectrum eigenstates become localized on the left end while the low-spectrum eigenstates are only slightly affected.
This non-Hermiticity induced localization on the left end in the mid-spectrum eigenstates suggests many-body NHSE, and the sharp contrast between the low- and mid-spectrum regimes suggests that the thermal eigenstates~(mid-spectrum) are much more vulnerable to the many-body NHSE than the MBL eigenstates~(low-spectrum).

The same behavior is observed in the $g$-dependence in the energy resolved half-chain R\'enyi entropies $S_2(L/2,L)$~(see Fig.~\ref{fig:2}(b)). 
At $g=0$ in the left panel, the low- and mid-spectrum eigenstates exhibit area-law and volume-law ~\cite{Bauer2013} entanglement scalings, respectively, as expected from the MBL and thermal phases.
In the presence of non-Hermiticity $g>0$ at different degrees (see the right panel), the half-chain entanglement of mid-spectrum eigenstates decreases drastically with $g$, showing that the many-body eigenstates become localized due to many-body NHSE. 
In contrast, the low-energy half-chain entanglement entropies remain resilient within the range shown here, and will not be significantly affected by NHSE until an extremely large $g$~\cite{SupplementaryMaterial}.

Finally, we show that the subsystem size $l$-dependence of the R\'enyi entropies $S_2(l,L)$ is in fact a more sensitive diagnostic for the influence of many-body NHSE in the thermal and MBL energy regimes at intermediate non-Hermiticity $g$ (see Fig.~\ref{fig:2}(c)). 
At $g=0$, the representative MBL and thermal states in the left and right panels, respectively, show  an area-law scaling and a universal volume-law behavior~\cite{Nakagawa2018}.
At small $g>0$, it is evident that the thermal state already shows a significant change in $S_2(l,L)$ at a small $g$, while the $S_2(l,L)$ of the MBL state remain unchanged.
Nonetheless, both thermal and MBL states manifest NHSE at a strong non-Hermiticity $g=0.7$, where $S_2(l,L)$ are asymmetric and peak at $l/L=N/L=1/6$.
This is expected for NHSE because all particles are accumulated on the left end at large enough $g$ regardless of the energy regime, where the nonzero entanglement can only build up for subsystem size $l/L$ in the vicinity of $N/L$~\cite{SupplementaryMaterial}.
The filling $N/L$-controlled asymmetric behavior in the subsystem-size $l$ dependence of R\'enyi entropy $S_2(l,L)$ therefore serves as a sensitive diagnostic for the presence of many-body NHSE.

{\textit{Dynamical properties and experimental observables.}}---
We now examine whether the sharp contrast between the responses of NHSE on the MBL and thermal states manifests in the dynamical properties. 
To examine the fate of MBL in our non-Hermitian GAA model, we examine the hallmark of MBL---initial state memory retention.
The time evolution is computed by the quantum trajectory formalism under no-jump condition~\cite{Daley2014}, which corresponds to continuously measured systems or post-selection. 
We consider product states in an $L=18$ open chain with $N=3$ particles as the initial states $|\psi_0\rangle$ to study their quench dynamics.
At time $t$, the state is given by $|\psi_t\rangle=e^{-i H t}|\psi_0\rangle/\sqrt{\langle\psi_0|e^{i H^\dagger t}e^{-i H t}|\psi_0\rangle}$. 
The non-Hermiticity not only affects the formalism of time evolution but also results in the energy non-conservation process. 
Nonetheless, we can still characterize different initial states with their steady-state energies at long time~\cite{SupplementaryMaterial}.

\begin{figure}[!t]
\centering{}
\includegraphics{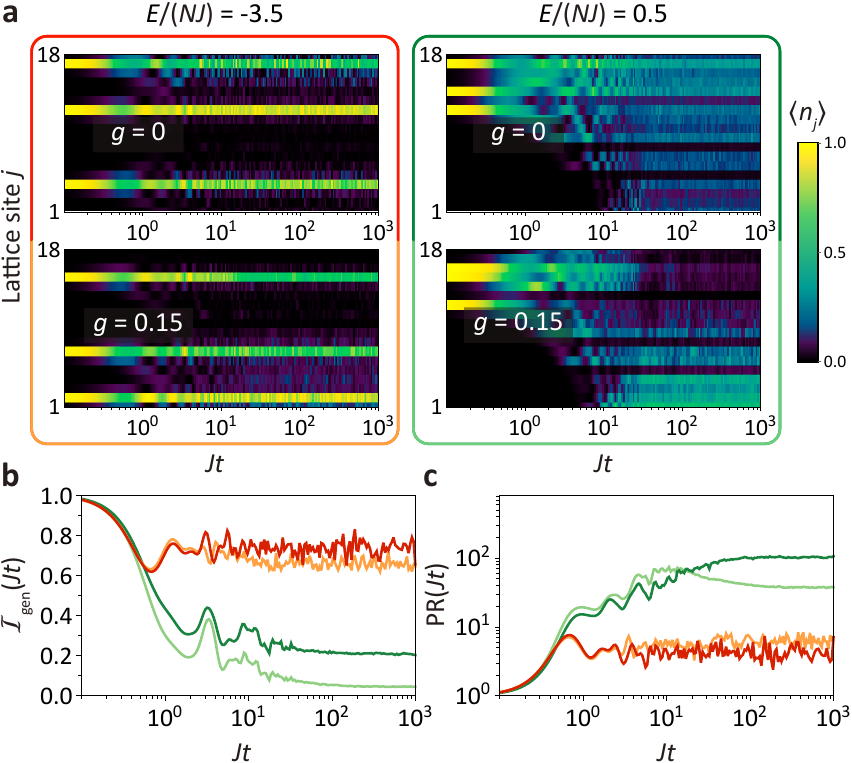}
\caption{\label{fig:3}
(a) Site-resolved dynamics for some initial product states with low~(left) and high~(right) real parts of energies at long-time.
The top and bottom panels correspond to the Hermitian and non-Hermitian cases, respectively.
To pick up these representative states, the global phases $\phi+\pi q (L+1)$ are chosen as left: $1.9\pi$~(top) and $0.4\pi$~(bottom) and right: $0.7\pi$~(top) and $0.7\pi$~(bottom).
(b,c)~Dynamics of (b)~generalized imbalance $\mathcal{I}_\text{gen}$ (c)~participation ratio $\text{PR}$ at low-~(orange) and mid-~(green) spectrum regimes in the Hermitian~(dark) and non-Hermitian~(light), which are used to diagnose initial-state memory and spread of information of initial state.
The parameters used in each curve follow the corresponding color of the boundary line in (a).
}
\end{figure}

We start by choosing the initial product states whose real parts of long-time energies are in the low- and mid-spectrum regimes, where the former and latter are considered to be in the MBL and thermal phases at $g=0$.
We show the time evolution of local density for four representative states in Fig.~\ref{fig:3}(a).
To examine the role of $g$ on a specific energy regime, we consider different $\phi$ and different initial state to maintain the same long-time energy.
For the low-energy state~(left), the persisting memory of the initial state at $g=0$~(top) breaks ergodicity and is consistent with states in the MBL phase.
A similar behavior is observed at $g=0.15$~(bottom), where the influence of NHSE can only be witnessed at a larger $g$.
In the mid-spectrum regime~(right) on the contrary, the initial product state rapidly spreads over the whole lattice at $g=0$~(top), reflecting the absence of initial-state memory.
At $g=0.15$~(bottom), although the information of the initial state is lost within short time, the long-time dynamics exhibits a clear accumulation of particles at the left end, which manifests the many-body NHSE.
This lost of the initial information is a sharp difference between the features of many-body NHSE and MBL.

The preservation of the information of initial state can be better expressed as the generalized imbalance~\cite{G2021,Guo2021,Morong2021}
\begin{equation}
\mathcal{I}_\text{gen}=\sum_{j=1}^{L}\beta_j n_j,
\end{equation}
where $\beta_j=N^{-1}$ and $-(L-N)^{-1}$ for initially occupied and unoccupied sites when $N<L/2$, respectively.
This ensures that all initial product states $|\psi_0\rangle$ correspond to unity generalized imbalance, i.e., $\langle\psi_0|\mathcal{I}_\text{gen}|\psi_0\rangle=1$.
For a thermal state, $\langle\psi_t|\mathcal{I}_\text{gen}|\psi_t\rangle$ would relax to zero at long time due to the same probability for a fermion to occupy all sites.
Since the perfect preservation of the information of the initial state is manifested by $\langle\psi_t|\mathcal{I}_\text{gen}|\psi_t\rangle\to 1$, the value of $\mathcal{I}_\text{gen}$ that roughly separates the MBL phases from other phases can be chosen as the case with equal probability for a particle to occupy the initially occupied and unoccupied sites, which corresponds to $\mathcal{I}_\text{gen}=(N/2)N^{-1}-(N/2)(L-N)^{-1}$.
In our case, the $N/L=1/6$ filling corresponds to $\mathcal{I}_\text{gen}=0.4$.
Therefore, the low-spectrum regime exhibits the initial-state memory retention at long time, while the relaxations of generalized imbalances in the mid-spectrum regime show the loss of information of the initial state~[Fig.~\ref{fig:3}(b)].
This means that the generalized imbalance can indicate the MBL phase.

We note that even though there is localization of particles induced by NHSE in the mid-spectrum regime, it does not preserve the information of the initial state, such that the localization induced by the quasiperiodic potential and NHSE are fundamentally different.
To show that NHSE can also result in the localization in the Fock space from the dynamics, we move to the participation ratio $\text{PR}(Jt)=\big[\sum_{n=1}^{\mathcal{D}_{N,L}}|\langle n|\psi_t\rangle|^4\big]^{-1}$ that directly quantifies the distribution of a many-body state in the Fock space.
Here $|n\rangle$ is a product state and $\mathcal{D}_{N,L}=L!/[(L-N)!N!]$ is the dimension of Hilbert space in the $N$ particle sector.
For a uniformly distributed state, the PR corresponds to an upper bound $\mathcal{D}_{N,L}$, while a localized state has a relatively small PR.
The long-time participation ratios shown in Fig.~\ref{fig:3}(c) directly reveal that both quasiperiodic potential and NHSE can suppress PR, meaning that NHSE can prevent a state from traversing the whole Hilbert space.

\begin{figure}[!t]
\centering{}
\includegraphics{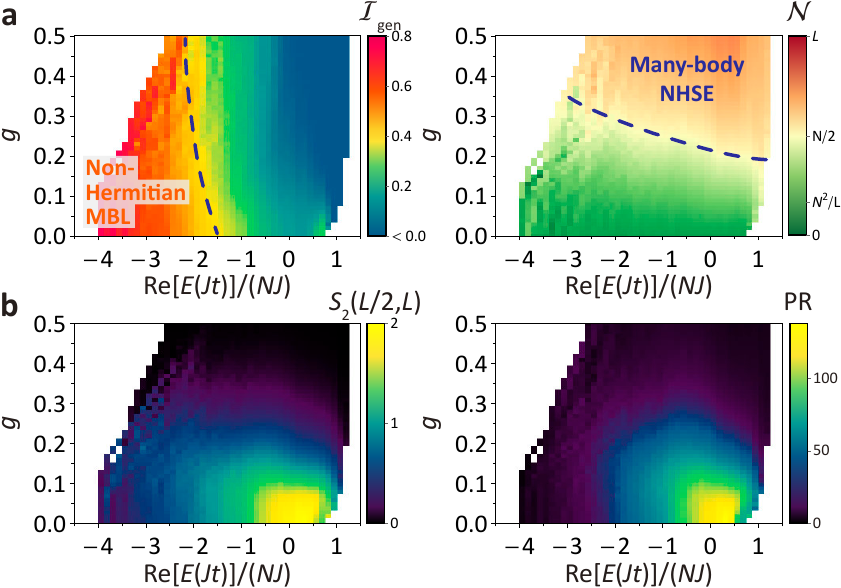}
\caption{\label{fig:4}
Snapshots of (a)~generalized imbalance $\mathcal{I}_\text{gen}$~(left) and number of particle at the first $N$ sites $\mathcal{N}$~(right) and (b)~half-chain R\'enyi entropy $S_2(L/2,L)$~(left) and participation ratio $\text{PR}$~(right) at $Jt=\text{1000}$. $L=18$ is considered in these plots.
}
\end{figure}

{\textit{The phase diagrams of MBL and many-body NHSE.}}---With these long-time behaviors of quench dynamics of initial product state, we can establish the phase diagrams of MBL and many-body NHSE in Fig.~\ref{fig:4} by a set of experimentally feasible quantities.
In the left panel of Fig.~\ref{fig:4}(a), we present the generalized imbalance at $Jt=\text{1000}$ as a function of the real part of energy at the same time.
The MBL phase is characterized by the parameter region with $\mathcal{I}_\text{gen}>0.4$, which is shown in red with dashed blue line as a guide to the eye, meaning that the MBL persists in the non-Hermitian system.
These results suggest that the parameter region of the non-Hermitian MBL phase would be slightly affected by the non-Hermiticity $g$, which can not be revealed by the level statistics.

To characterize the NHSE, we calculate the total particle numbers at the first $N$ sites $\mathcal{N}(Jt)=\langle\psi_t|\sum_{j=1}^{N}n_j|\psi_t\rangle$ at long time.
In the absence of NHSE, $\mathcal{N}$ is expected to be $N^2/L$, which is reflected in the horizontal cut of the right panel of Fig.~\ref{fig:4}(a) at $g=0$.
In the strong $g$ limit, $\mathcal{N}$ would approach $N$, such that $\mathcal{N}=N/2$~(indicating by a dashed blue line) is a good reference value to characterize NHSE.
It is evident that the states with larger $\text{Re}[E(Jt)]/(NJ)$ are easier to be affected by NHSE, which coincides with the entanglement reduction induced by NHSE in Fig.~\ref{fig:2}(b).
This result again reflects the fact that MBL phase can escape NHSE, and we note that both extended and localized phases are dominated by NHSE at an extremely large $g$.

The entanglement reduction in the mid-spectrum regime is also observed in the left panel of Fig.~\ref{fig:4}(b), which presents the half-chain R\'enyi entropy at long time.
Besides, the MBL also results in the low entanglement entropy.
This is because the many-body states here become more localized in the Fock space.
To reveal this correspondence between the entanglement reduction and the localization in the Fock space, we consider the long-time participation ratio in the right panel of Fig.~\ref{fig:4}(b), which shows the similar behavior as that of the R\'enyi entropy.
Accordingly, we can attribute two localized phases with small participation ratios to MBL and many-body NHSE by the large long-time generalized imbalance and the accumulation of particles at the open boundary, respectively, from which the phase diagram in Fig.~\ref{fig:1}(b) is established.

{\textit{Conclusion.}}---We have shown that NHSE has a significant impact on the thermal states, while the MBL states require strong non-Hermiticity to be altered.
These features are manifested in both the many-body eigenstates and the quench dynamics of product states.
We note that here we focus on distinct behaviors of individual phases, while their separations and intermediate stages require further studies.
Based on the experimentally feasible quantities we provided, we expect our analysis can be applied to generic interacting non-Hermitian systems with random disorder or quasiperiodic potential.

{\textit{Acknowledgement.}}---
We thank Qian-Rui Huang and Po-Jen Hsu for technical supports.
Y.-C.W., K.S., and H.H.J. acknowledge support from the Ministry of Science and Technology~(MOST), Taiwan, under the Grant No. MOST-109-2112-M-001-035-MY3.
Y.-C.W. and J.-S.Y. are supported by the Ministry of Science and Technology, Taiwan~(Grant No. MOST-110-2112-M-003-008-MY3).
H.H.J. and J.-S.Y. are also grateful for support from National Center for Theoretical Sciences in Taiwan.

\bibliography{NHSE_on_MBL_and_thermal_phases_v2}
\end{document}